\documentclass{article}
\usepackage{lineno}
\usepackage{xcolor}
\usepackage{amsmath}

\usepackage{PRIMEarxiv}

\usepackage[utf8]{inputenc} 
\usepackage[T1]{fontenc}    
\usepackage{hyperref}       
\usepackage{url}            
\usepackage{booktabs}       
\usepackage{amsfonts}       
\usepackage{microtype}      
\usepackage{lipsum}
\usepackage{fancyhdr}       
\usepackage{graphicx}       
\usepackage{lineno}
\usepackage{nicefrac, xfrac}

\usepackage{authblk}
\usepackage{parskip}
\usepackage[absolute]{textpos}

\title{Development of a 100 ps TDC based on a Kintex 7 FPGA for the High Granular Neutron Time-of-Flight detector for the BM@N experiment}

\author[*,a]{D. Finogeev}
\author[a,b]{F. Guber}
\author[a]{A. Izvestnyy}
\author[a]{N. Karpushkin}
\author[a,b]{A. Makhnev}
\author[a]{S. Morozov}
\author[a]{D. Serebryakov}
\affil[a]{Institute for Nuclear Research of the Russian Academy of Sciences, Moscow, Russia}
\affil[b]{Moscow Institute of Physics and Technology, Dolgoprudny, Russia}

\begin{document}
\maketitle

\begin{abstract}
The prototype of a TDC board has been developed for the new high granular time-of-flight neutron detector (HGND). The board is based on the standard LVDS 4x asynchronous oversampling using the xc7k160 FPGA with a 100~ps bin width. The HGND is being developed for the BM@N (Baryonic Matter at Nuclotron) experiment to identify neutrons and to measure their energies in heavy-ion collisions at ion beam energies up to 4~A~GeV. The HGND consists of about 2000 scintillator detectors (cells) with a size of $40~\times~40~\times~25~mm^3$ and light readout with EQR15 11-6060D-S photodetectors. To measure the time resolution of the scintillator cells, the two-channel FPGA TDC board prototype with two scintillator cells was tested with an electron beam of “Pakhra” synchrotron at the LPI institute (Moscow, Russia). The measured cell time resolution is 146 ps, which is in a good agreement with the 142~ps time resolution measured with a 12-bit~@~5~GS/s CAEN DT5742 digitizer. For the full HGND, the TDC readout board with three such FPGAs will read 250 channels. In total, eight such TDC boards will be used for the full HGND at the BM@N experiment.

\end{abstract}

\begin{textblock*}{\textwidth}(2.5cm,26cm)
  \begin{flushleft}
    \noindent\rule{7cm}{0.4pt} \\
    * Corresponding author. Email: finogeev@inr.ru
  \end{flushleft}
\end{textblock*}


\section{Introduction}
\label{sec:intro}

The physics program of the BM@N fixed target experiment~\cite{Kapishin:2020cwk, Senger:2022bzm} aims at studying the Equation of State (EoS) of high-density nuclear matter created in heavy-ion collisions at beam energies up to 4~A~GeV. It is anticipated that neutron flow will be a sensitive probe to study the isospin component of the equation of state (EoS) of high-density nuclear matter. To measure the neutron flow, the new high granular neutron time-of-flight detector (HGND) is under development at INR RAS, Moscow \cite{Guber:2023jxf}. The HGND will identify high-energy neutrons (up to 4~GeV) and measure their energies by the time of flight. The estimated energy resolution is 15\%, and the neutron detection efficiency is about 60\% for the 4~GeV neutrons. \par

The HGND consists of about 2000 plastic scintillator detectors (cells) with a size of $40~\times~40~\times~25~mm^3$. The SiPM (silicon photomultiplier) EQR15~11-6060D-S~\cite{dtsh:eqr}, with a sensitive area of $6~\times~6~mm^2$, is used for light readout from each cell. A JINR-produced scintillator based on polystyrene with additions of 1.5\% paraterphenyl and 0.01\% POPOP is considered as the main option. The time resolution of cells made of fast scintillators equipped with EQR15~11-6060D-S is about 120~ps~\cite{Guber:2023ktn}. Due to the high total number of electronics channels of the HGND, a multi-channel readout system needs to be developed. The digital part of the readout is designed using a Kintex FPGA (Field Programmable Gate Array) chip, which incorporates a TDC (Time to Digital Converter) with a bin width of 100~ps. The TDC is realized to perform precise timestamp and amplitude measurements using the Time-over-Threshold (ToT) method. The FPGA-based readout has time synchronization with the White Rabbit~\cite{Serrano:wrp} network and will be integrated into the BM@N common DAQ (data acquisition system).

\section{TDC implementation in the FPGA}
\label{sec:tdc_imp}

\begin{figure}[h]
\centering
\includegraphics[width=.8\textwidth]{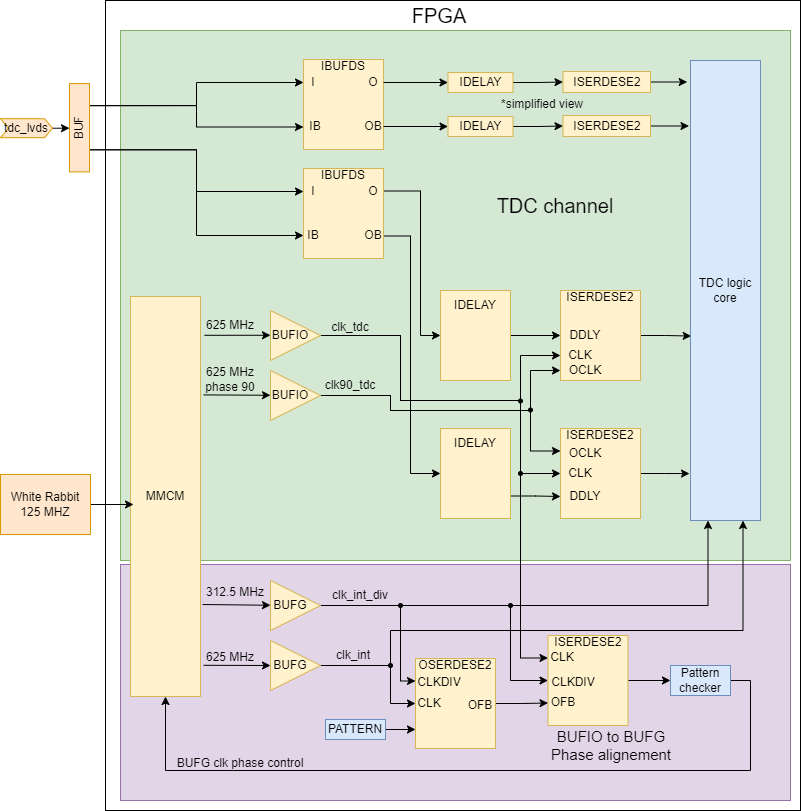}
\caption{\label{fig:tdc_line} FPGA TDC clock scheme for a single channel. The TDC channel, represented by the green (top) section, includes IDELAY and ISERDESE2 primitives, and this section is replicated for each TDC input channel. The pink (bottom) section represents the BUFIO to BUFG phase alignment, which must be in each FPGA bank.}
\end{figure}

The TDC is based on the Kintex-7 input serial-to-parallel converter with oversampling capability and programmable delay. The design is based on Xilinx recommendations~\cite{XAPP523}, and uses only documented features of the FPGA within its specifications. The simplified TDC schematic is presented in figure~\ref{fig:tdc_line}. The TDC channel section (green, top) is replicated for every TDC input channel. The input TDC lines are connected to the input differential FPGA buffers (IBUFDS), which give four TDC input lines. Each TDC input line is delayed using its own IDELAY core. The delayed signal for each TDC line is captured four times within a single period at a frequency of 625 MHz using the ISERDESE2 (serial-to-parallel converter) core. The outputs of the four ISERDESE2 cores are digitally processed in the TDC logic core to perform time calculations.

The FPGA ISERDESE2 core is driven with a 625 MHz clock. This converter has a so-called “oversample” mode, where the input signal is sampled four times during each clock period. Sampling is performed with two clocks, precisely shifted by $90^\circ$. The ISERDES2 logic has three flip-flops in each sampling channel to minimize the metastability and to synchronize the samples at the output stage to one clock edge. The clocking of the flip-flops ensures that the time between sampling of the two connected flip-flops is not less than \nicefrac{3}{4} of the clock period. Sampling the TDC input four times during each clock period at 625~MHz using ISERDESE2 results in a sampling interval of 400~ps.\par

Four 400 ps resolution channels, driven by the same clock, are time-shifted by 100, 200, and 300 ps from the first one. They sample the shifted input signals simultaneously, enabling time measurement with a 100 ps bin width. These channels utilize a programmable delay core (IDELAY) to achieve the time shifts. Two differential input pairs (4 pins of the FPGA) are used to build one TDC channel. The LVCP22~\cite{LVCP22} LVDS crosspoint switch is employed for distributing the TDC channel signals to two FPGA LVDS inputs. Additionally, it switches every TDC channel to the FPGA pulse generator in order to facilitate calibration tasks. With a Peak-to-Peak Jitter of 50 ps, the LVCP22 guarantees sufficient accuracy for the 100 ps TDC design. In the ultimate TDC design, the LVDS pair will directly connect to two FPGA LVDS inputs by utilizing just one output buffer from the crosspoint switch. This configuration is aimed at minimizing timing skew between the TDC channel inputs. Precise sampling clocks are provided by a special clock buffer, BUFIO.\par

In Figure~\ref{fig:tdc_line}, the pink (bottom) section illustrates the implementation of the BUFIO to BUFG phase alignment, which must be in each FPGA bank. This component aligns clocks from BUFIO and BUFG buffers. Specifically, a fixed data pattern is transmitted by OSERDESE2 using the BUFG clock and received by ISERDESE2 using the BUFIO clock. By comparing the transmitted and received patterns, the BUFIO to BUFG phase alignment is performed.

Two clock groups are managed by the Multi-Mode Clock Manager (MMCME2) core. The ISERDESE2 uses clocks from BUFIO. The ISERDESE2 outputs are then sampled by the fabric (general-purpose) flip-flops of the TDC logic core. The fabric flip-flops are driven by the fabric clock, which is supplied by a global clock buffer (BUFG). There is no way to calculate the real phase shift between fabric clock signals and BUFIO clocks. Therefore, to set the required phase shift between two groups of clocks, one pair of parallel-to-serial converter (OSERDESE2), driven by BUFG, and a serial-to-parallel converter driven by BUFIO clocks is used. The phase alignment is performed by scanning the phase of the global clock while comparing OSERDESE2 output pattern with ISERDESE2 input pattern. The goal is to set  the phase of the global clock in the middle point between two unstable phase ranges. Each FPGA bank has its own phase alignment  because the BUFIO network is local in each bank. All MMCME2 cores use a single 125 MHz source clock provided by the White Rabbit system synchronous to the experiment data acquisition, which allows absolute time measurements in all detectors in the BM@N experiment.\par

\begin{figure}[h]
\centering
\includegraphics[width=.8\textwidth]{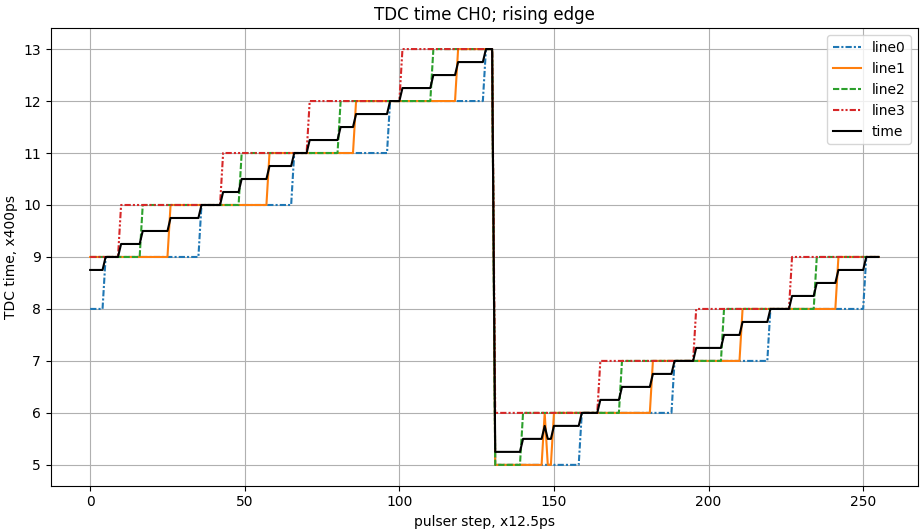}
\caption{\label{fig:tdc_time_scan} Four TDC lines and the resulting time dependence on the pulse time shift are shown. Pulses are generated by FPGA MMCM synchronously to the TDC clock with a phase step of 12.5ps. The single scan pass was taken with a digital FPGA logic analyzer.}
\end{figure}

In order to decrease a high processing frequency (625 MHz), two 4-bit samples from a serial-to-parallel converter are composed into an 8-bit vector at half of the initial frequency. The vector’s pattern represents the last 3.2 ns of the input signal. The outputs of the 4 sampling circuits are connected to the TDC logic core, which performs the final time calculations at a 312.5 MHz clock frequency. In the vector, the first non-zero bit denotes the occurrence of the rising edge of the TDC signal, while the last non-zero bit indicates the falling edge. Within the logic core, four patterns are compared with each other. Each pattern corresponds to a TDC line that has been time-shifted by 100 ps relative to one another. The specific combination of patterns depends on the time difference between the rising edge of the TDC signal and the rising edge of the ISERDESE2 clock. When the time difference is less than 100 ps, only one pattern from a single line exhibits the signal. If the time difference falls between 100 and 200 ps, two patterns are involved, and so on. The sampling results from all four circuits are shown in the figure~\ref{fig:tdc_time_scan}.\par

The minimum detected pulse length for the current version of the TDC logic is 6.4~ns, and the dead time is 9.6~ns. These values can be reduced to 1.6~ns each if needed. The length of a signal registered from the HGND cell, corresponding to the Minimum Ionization Particle (MIP), is 150 ns. To prevent oscillation on the threshold comparator, a veto is applied for 32 ns after the falling edge of the signal. The estimated HGND channel load is less than 100 kHz. The current time limitations of the TDC logic core are derived from the simplicity of the selected algorithm, and the performance of HGND will not be constrained by the TDC readout.\par

The maximum number of TDC channels that can be implemented in an FPGA is equal to the total number of available LVDS inputs and input logic (ILOGIC) resources. The Kintex 7 FPGA single bank allows to create 12 TDC channels: 24 LVDS input lines and 48 ILOGIC blocks of 50 available. 
One additional ISERDES2 - OSERDES2 pair is required in each FPGA bank for phase alignment. The xc7k160tffg676 FPGA will be used for the HGND detector readout board. In the forthcoming configuration, a total of seven out of the eight available FPGA banks will be allocated to accommodate 84 TDC channels per FPGA. The single readout board with three FPGAs will hold 250 channels. In total, eight such boards will provide all 2000 detector channels of the HGND.

\section{TDC time alignment}
\label{sec:tdc_alignment}

\begin{figure}[h!]
\centering
\includegraphics[width=0.8\textwidth]{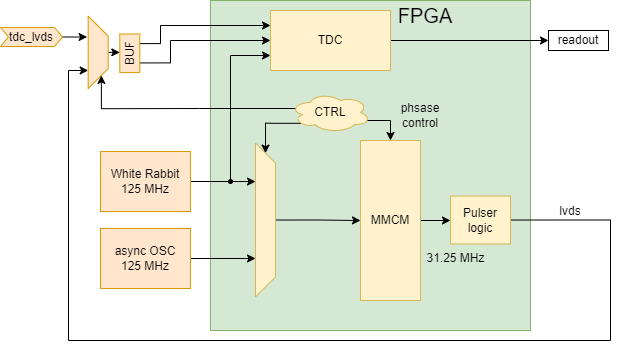}
\caption{\label{fig:tdc_pulser} The scheme of TDC calibration pulse generator.}
\end{figure}

The input delay (IDELAY) FPGA primitive provides controllable delay for each TDC sampling circuit. The IDELAY block is controlled by the DLL (delay locked loop) in the IDELAYCTRL primitive. Each IDELAY block has a delay range from 0 to 31 steps. The DLL control clock, with a frequency of 402.7~MHz, sets the delay step value to 38.8~ps. Thus, it ensures the closest to 100~ps TDC relative line delays: 104  ($13\times38.8 - 400$), 194 ($5\times38.8$) and 298 ($18\times38.8 - 400$)~ps, available in the FPGA PLL (phase locked loop) generated from the source clock of 125~MHz. The TDC alignment procedure is performed using an external LVCP22 crosspoint switch, which connects each TDC channel to the pulse generator. The pulse generator is based on the FPGA MMCM block, and it can generate pulses synchronously or asynchronously to the TDC clock with 12.5~ps/step precision phase shift. The schematic of the pulse generator for TDC calibration is presented in figure~\ref{fig:tdc_pulser}.\par

\begin{figure}[h!]
\centering
\includegraphics[width=0.8\textwidth]{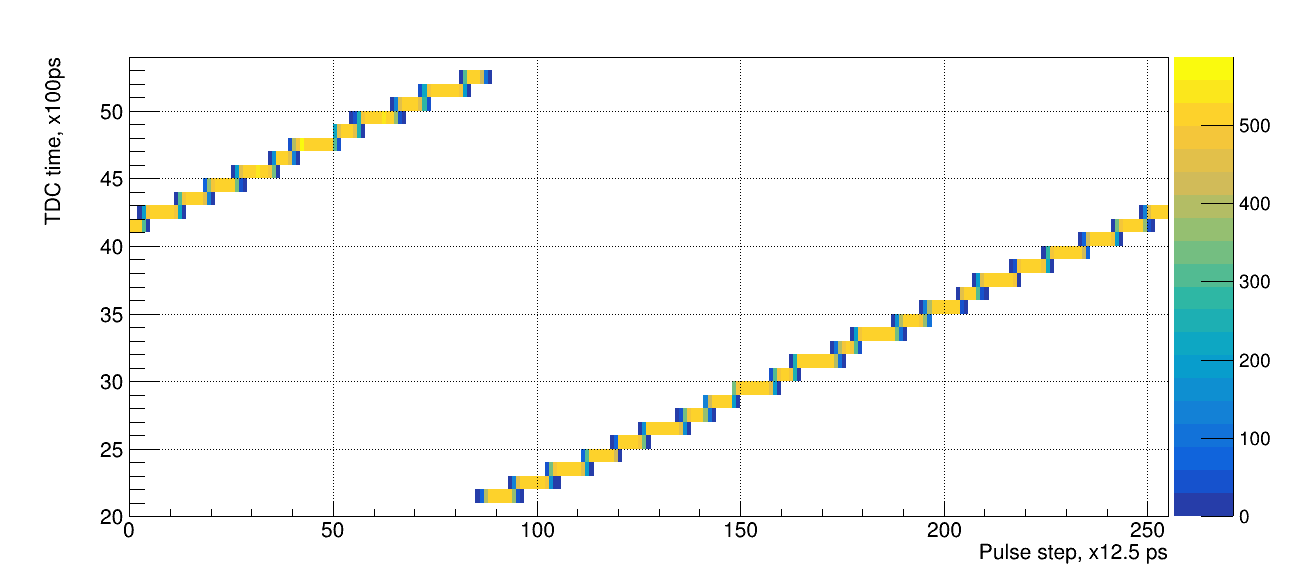}
\caption{\label{fig:tdc_calib_scan} The TDC time dependence on pulse time shift. Pulses are generated by FPGA MMCM synchronous to TDC clock, and the time step is 12.5~ps. Data was taken with PC readout, 1000 events per single time shift step.}
\end{figure}

The main goal of the calibration procedure is to provide a 100~ps relative TDC line alignment with a 200~ps shift in each LVDS pair. The delays between the TDC lines, which are higher than 400~ps, are aligned digitally in the logic. The TDC time scan with pulses synchronous to the TDC clock with a 12.5~ps step is presented in the figure~\ref{fig:tdc_calib_scan}. Such a synchronous time scan is used for the differential nonlinearity correction performed in software in a semi-manual manner. The differential nonlinearity correction will be implemented later in FPGA and combined with the TDC delay calibration procedure.

\begin{figure}[h!]
\centering
\includegraphics[width=.45\textwidth]{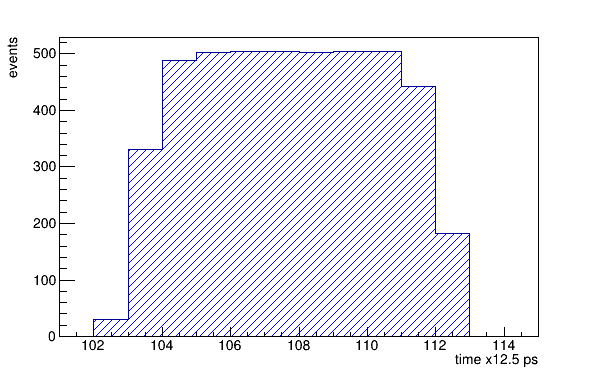}
\qquad
\includegraphics[width=.45\textwidth]{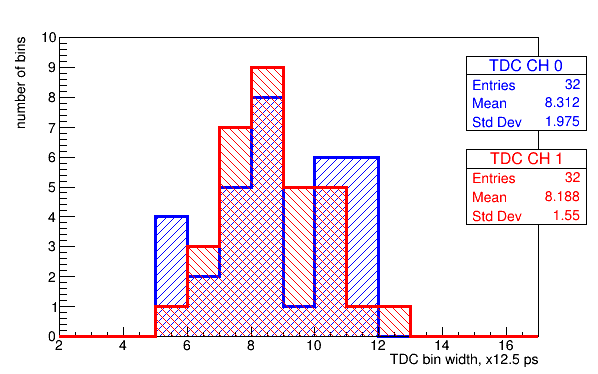}
\caption{\label{fig:bin_width_sync} The TDC bin width measured with pulses synchronous to the TDC clock. Left: Typical TDC bin profile taken with a synchronous time scan. Right: The TDC bin width distribution for both TDC channels. The bin width value taken at the amplitude level is 30\% in the bin profile.}
\end{figure}

\begin{figure}[h!]
\centering
\includegraphics[width=.45\textwidth]{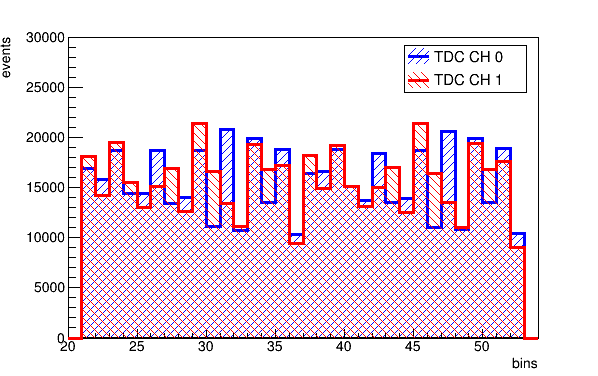}
\qquad
\includegraphics[width=.45\textwidth]{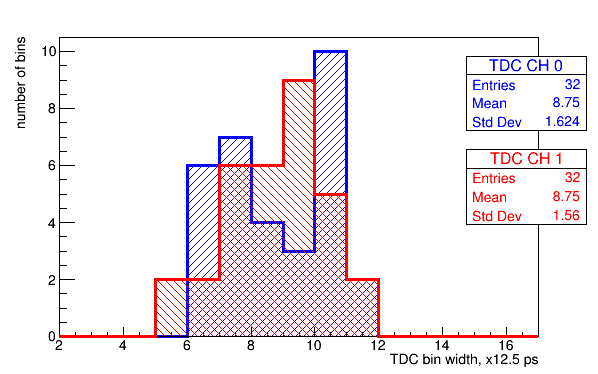}
\caption{\label{fig:bin_width_async} The TDC bin width is measured with pulses asynchronous to the TDC clock. Left: Asynchronous pulses distribution by TDC bins for both channels. Right: TDC bin width distribution calculated by considering that the probability to have an asynchronous pulse in the bin is directly proportional to the bin width.}
\end{figure}

The typical TDC bin profile is shown in figure~\ref{fig:bin_width_sync} (left). The distribution of bin width taken at the profile amplitude level of 30\% for both channels is presented in figure~\ref{fig:bin_width_sync} (right). The TDC bin width measured with pulses asynchronous to the TDC clock (code density test~\cite{Machado:2019rkz}) is presented in figure~\ref{fig:bin_width_async}. The number of measured events in each TDC bin from pulses asynchronous to the TDC clock is shown in the figure~\ref{fig:bin_width_async} (left). Considering that the probability to have an asynchronous pulse in the bin is directly proportional to the bin width, the bin width can be calculated with the equation \ref{eq:bin_width}:

\begin{equation}\label{eq:bin_width}
  \begin{split}
    {\tau}_{bin} = \frac{{N_{bin}}{M}}{N_{tot}}\cdot{100ps}
  \end{split}
\end{equation}

\noindent where ${\tau}_{bin}$ is the bin width, ${N_{bin}}$ is the number of events collected in the bin, ${M}$ is the total number of bins, and ${N_{tot}}$ is the total number of events in all bins. If all bins have the same number of events then all bin widths will be set to 100ps. Figure~\ref{fig:bin_width_async} (right) shows a calculated TDC bin distribution. The method based on bin profile taken with synchronous pulses has the same result as the method of code density test taken with asynchronous pulses. Both measurements show ${1.6}\times{12.5} = 20~ps$ RMS of the TDC bin width distribution. The main contribution to the TDC bin width distribution is the TDC line delay alignment step of 38.8~ps.\par

\section{TDC and readout test}
\label{sec:tdc_test}

\begin{figure}[h!]
\centering
\includegraphics[width=0.5\textwidth]{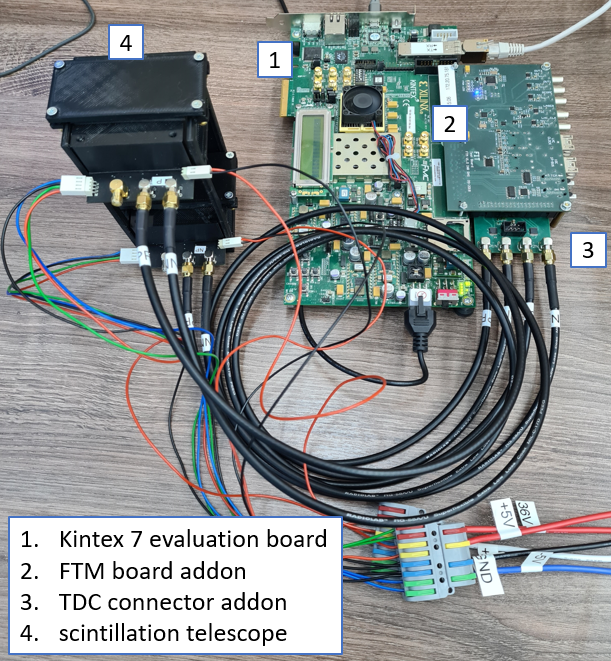}
\caption{\label{fig:hodo_photo} The photo of the test setup.}
\end{figure}

The Xilinx Evaluation board KC705 has been used for the FPGA TDC firmware developments. The FTM (FIT Test Module) board~\cite{Finogeev:2020qkf} with the TDC addon is connected to the FMC (FPGA Mezzanine Card) connectors. The TDC addon board holds SMA connectors and the LVDS line multiplexers. The FPGA evaluation board sends the data to the PC via Ethernet using the IPbus FPGA core~\cite{GhabrousLarrea:2015yfx}. In order to test the TDC performance, the telescope setup consisting of the two scintillator detectors (cells) has been used. The cells of size $40~\times~40~\times~25~mm^3$ are placed at a distance of 10~cm. Each cell contains the SiPM EQR15~11-6060D-S coupled with a scintillator produced at JINR. The SiPM is soldered on a PCB with the preamplifier and the threshold comparator ADCMP553 (125~ps overdrive dispersion,~\cite{ADCMP553}) with LVDS output. Each channel of an LVDS line is splitted on the TDC addon to provide 4 single TDC lines. The photo of the test setup is shown in figure~\ref{fig:hodo_photo}.

\begin{figure}[h!]
\centering
\includegraphics[width=.45\textwidth]{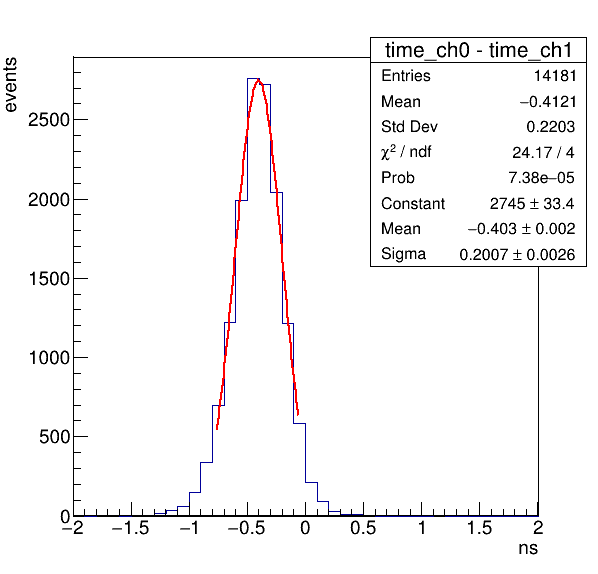}
\qquad
\includegraphics[width=.45\textwidth]{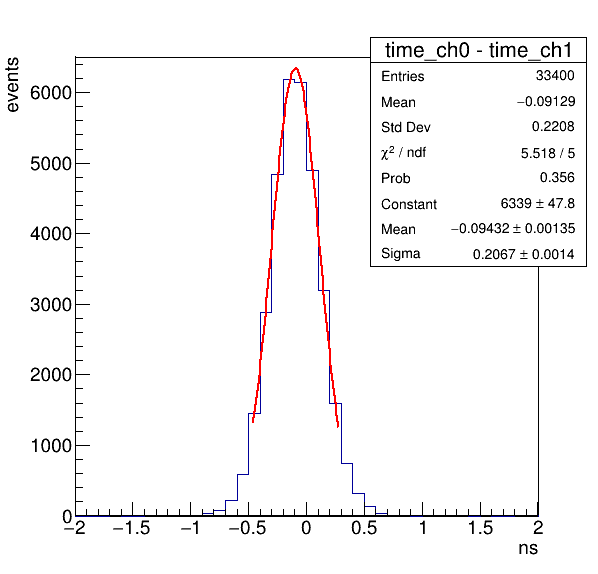}
\caption{\label{fig:time_res} The time difference distributions of two cells of the telescope measured with the CAEN digitizer (left) and the FPGA TDC prototype board (right).}
\end{figure}

The time resolution measurements with the FPGA TDC prototype board were performed with the 280~MeV electron beam on the “Pakhra” synchrotron in LPI (Moscow, Russia). The time resolution was also measured using a 12-bit~@~5~GS/s CAEN DT5742 digitizer for readout. Figure~\ref{fig:time_res} shows the time difference distribution of two cells measured with a CAEN digitizer (left) and the FPGA TDC prototype board (right). The single cell time resolution is $201 / \sqrt{2} = 142~ps$ (CAEN) and  $207 / \sqrt{2} = 146~ps$ (TDC).

\begin{figure}[h!]
\centering
\includegraphics[width=.7\textwidth]{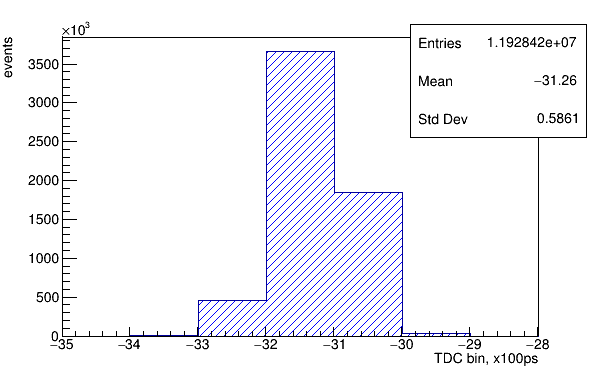}
\caption{\label{fig:tdc_gen_res} The time difference distribution between two FPGA TDC channels measured with the data generator DG2040 (Cycle-to-Cycle Jitter 5ps).}
\end{figure}

The time resolution of the FPGA TDC was measured with the data generator DG2040 (Cycle-to-Cycle Jitter 5ps). The delay scan between the two FPGA TDC channels was done with a 10 ps step in the range from -3 to 3~ns. The corrected time difference distribution  according to the results of this scan is shown in figure~\ref{fig:tdc_gen_res}. The RMS per channel is $59 / \sqrt{2} = 42~ps$, which demonstrates good agreement with the estimated quantization error typically obtained using ${\tau}_{bin} / \sqrt{6} = 40.8~ps$~\cite{Szplet:2019pp}.

\section{Conclusions}
\label{sec:concl}

The FPGA TDC with a 100~ps bin width based on the Kintex-7 input serial-to-parallel converter with oversampling capability and programmable delay has been developed. The TDC design was developed with Xilinx recommendations, and it uses only documented features of the FPGA within its specifications. One of the benefits of using TDC based on standard FPGA specification and application notes is the guaranteed system stability. The stability of the design is verified using the software suite produced by the FPGA manufacturer. The time resolution of the TDC channel measured with the data generator DG2040 (Cycle-to-Cycle Jitter 5~ps) is 42~ps. The developed TDC is based on the asynchronous oversampling with the FPGA primitive ISERDESE2. It shows good results in terms of time resolution and FPGA resource utilization, in comparison to the multi-phase clock sampling architecture of the FPGA TDC design~\cite{Cojocariu:2023kuu}. The  FPGA TDC board prototype was tested with scintillator detectors on the electron beam of the “Pakhra” synchrotron in LPI (Moscow, Russia). The measured time resolution of the scintillator detector with the FPGA TDC board prototype is 146 ps, which is in good agreement with the time resolution measured with a 12-bit~@~5~GS/s CAEN DT5742 digitizer (142~ps). The developed TDC based on the xc7k160tffg676 FPGA will be used in the HGND readout boards for the BM@N experiment located on the Nuclotron extracted ion beams at JINR, Dubna, Russia.

\section{Acknowledgements}
\label{sec:ackn}

This work was carried out at the Institute for Nuclear Research, Russian Academy of Sciences, and supported by the Russian Scientific Foundation grant №22-12-00132.


\begin{thebibliography}{10}

\bibitem{Kapishin:2020cwk}
Kapishin M., \emph{Heavy Ion BM@N and MPD Experiments at NICA}, {JPS Conf.Proc. 32 (2020) 010093}.

\bibitem{Senger:2022bzm}
BM@N Collaboration, Peter Senger, \emph{The heavy-ion program at the upgraded Baryonic Matter@Nuclotron Experiment at NICA}, {PoS CPOD2021 (2022) 033}.

\bibitem{Guber:2023jxf}
F.Guber et al., \emph{Development of High Granular Neutron Time-of-Flight Detector for the BM@N experiment}, {9 2023}, {arXiv: 2309.09610}.


\bibitem{dtsh:eqr}
Novel Device Laboratory, \emph{EQR15 Series SiPMs datasheet}, \url{http://www.ndl-sipm.net/PDF/Datasheet-EQR15.pdf}.


\bibitem{Guber:2023ktn}
F.~Guber, A.~Ivashkin, N.~Karpushkin, A.~Makhnev, S.~Morozov, D.~Serebryakov, V.~Baskov, and V.~Polyansky,
 \emph{Measurement of Time Resolution of Scintillation Detectors with EQR-15 Silicon Photodetectors for the Time-of-Flight Neutron Detector of the BM@N Experiment}, {9 2023}, {arXiv: 2309.03614}.

 
\bibitem{Serrano:wrp}
J. Serrano, P. Alvarez, M. Cattin, E. G. Cota, P. M. J. H. Lewis, T. Włostowski et al., \emph{The White Rabbit Project}, {Proceedings of ICALEPCS TUC004, Kobe, Japan, 2009}.

\bibitem{XAPP523}
Xilinx, Marc Defossez, \emph{XAPP523: LVDS 4x Asynchronous Oversampling Using 7 Series FPGAs and Zynq-7000 AP SoCs}.

\bibitem{LVCP22}
Texas Instuments, \emph{SN65LVCP22: 2x2 LVDS CROSSPOINT SWITCH datasheet}, \url{https://www.integrated-circuit.com/pdf/565/741.pdf}.

\bibitem{Machado:2019rkz}
Rui Machado, Jorge Cabral, Filipe Serra Alves, \emph{Recent Developments and Challenges in FPGA-Based Time-to-Digital Converters}, {IEEE Trans.Instrum.Measur. 68 (2019) 11, 4205-4221}.

\bibitem{Finogeev:2020qkf}
ALICE Collaboration, D. Finogeev, \emph{Readout system of the ALICE Fast Interaction Trigger}, {JINST 15 (2020) 09, C09005}.

\bibitem{GhabrousLarrea:2015yfx}
C. Ghabrous Larrea, K. Harder, D. Newbold, D. Sankey, A. Rose, A. Thea and T. Williams, \emph{IPbus: a flexible Ethernet-based control system for xTCA hardware}, {JINST 10 (2015) no.02, C02019}.

\bibitem{ADCMP553}
Analog Devices, \emph{ADCMP553: Single-Supply High Speed PECL/LVPECL Comparators datasheet}, \url{https://www.farnell.com/datasheets/1903306.pdf}.

\bibitem{Szplet:2019pp}
R. Szplet, R. Szymanowski, and D. Sondej, \emph{Measurement Uncertainty of Precise Interpolating Time Counters}, {IEEE Transactions on Instrumentation and Measurement (2019)}.


\bibitem{Cojocariu:2023kuu}
LHCb RICH Collaboration, L.N. Cojocariu, et al. \emph{A multi-channel TDC-in-FPGA with 150 ps bins for time-resolved readout of Cherenkov photons}, {Nucl.Instrum.Meth.A 1055 (2023) 168483}.

\end{thebibliography}
\end{document}